# Case-Level Verification in Scanner–LLM Cascades: Overcoming the Alert Aggregation Bottleneck to Expand the FRR–TPR Trade-off Space

1st Hao Sun
Shenzhen Information Security Management Center
1st Yibin Yao
Shenzhen Information Security Management Center
2nd Chaohai Xie
Shenzhen Secidea Network Security Technology Co. Ltd
3rd Yuqun Lin*
Shenzhen Information Security Management Center

**Abstract**

Dynamic Application Security Testing (DAST) scanners achieve high recall but also produce a large number of false positives, resulting in substantial manual triage costs. Large Language Models (LLMs), when used for independent detection, achieve extremely high recall (95.4%–100%) but also exhibit prohibitively high false positive rates (49.6%–85.0%), precluding their use as standalone replacements for scanners. A natural solution is a two-stage cascade consisting of scanner detection followed by LLM verification. However, a verification-granularity issue that has long been overlooked in practice creates a structural bottleneck: alert aggregation binds multiple true and false cases into a shared decision unit, such that removing a false positive inevitably eliminates true positives aggregated within the same alert group. This creates a trade-off bottleneck between the False-positive Reduction Rate (FRR) and the True-positive Rate (TPR). We formalize this bottleneck as Proposition $F_{alert} \subseteq F_{case}$ and introduce a Case-Level, per-case verification strategy that shifts the decision granularity from the alert level to the instance level, independently replaying HTTP requests and making an independent determination for each detected case. Evaluation on the dual testbeds of Damn Vulnerable Web Application (DVWA) and WebGoat shows that the empirically best Alert-Level operating point achieves FRR=42.86% (TPR=51.7%). The Case-Level Baseline achieves FRR=47.6%, an improvement of 4.7 percentage points (+4.7 pp), while the Case-Focused Evidence Verification Prompt (CEV-Prompt) increases TPR from 55.2% to 62.1% at the same FRR.

## 1 Introduction

### 1.1 Background and Motivation

Web applications constitute a core component of modern information systems. Dynamic Application Security Testing (DAST) scanners identify vulnerabilities such as SQL injection, cross-site scripting (XSS), command injection, and file inclusion by injecting crafted payloads into a target and observing the resulting responses. They are among the most widely adopted automated detection techniques in industry [1-3]. Nevertheless, DAST is characterized by both high recall and a high false-positive burden. Prior evaluations of vulnerability scanners have shown that both commercial and open-source scanners generally produce substantial proportions of false positives, making manual triage the most costly and least

* Corresponding author.
Hao Sun and Yibin Yao contributed equally to this work and should be considered co-first authors.

automatable stage of the vulnerability-management pipeline [2,4].

In recent years, Large Language Models (LLMs) have demonstrated substantial capabilities in code understanding, semantic reasoning, and security analysis, prompting their application to source-code vulnerability detection, log analysis, and automated penetration testing [5-8]. However, the FPR of standalone LLM-based detection is likewise high, preventing LLMs from replacing scanners independently. The two technologies are therefore better viewed as complementary: the scanner prioritizes comprehensive detection, whereas the LLM filters spurious findings.

### 1.2 Core Tension and Research Questions

A natural way to combine a scanner with an LLM is to construct a two-stage cascade in which the scanner emits alerts and the LLM verifies them individually to filter false positives. In engineering implementations, however, verification granularity has long been overlooked. Before raw scanner detections reach either manual or LLM-based review, they are often aggregated and deduplicated: dozens of independent detections targeting different URLs, parameters, and payloads are collapsed into only a few alerts. False positives and true positives (TPs) therefore coexist within the same decision unit.

When an LLM makes decisions at this aggregated granularity, each decision must apply to an entire group of cases despite substantial information loss, thereby creating an FRR–TPR trade-off bottleneck: removing an alert that contains a false positive inevitably eliminates true positives aggregated within the same alert group. This data-structural fact forms the point of departure for all subsequent analyses.

Accordingly, this study addresses two research questions:

- RQ1: What specific mechanism underlies the FRR–TPR trade-off bottleneck introduced by alert aggregation, and can it be formalized?
- RQ2: Can shifting verification granularity to the case level (by independently replaying and adjudicating each case) expand the set of feasible FRR–TPR operating points?

### 1.3 Main Contributions

The contributions of this work span methodological, empirical, and engineering dimensions. Methodologically, we design and systematically evaluate a Case-Level, per-case verification strategy. By shifting verification granularity from a shared alert-level decision to independent replay and adjudication of individual cases, the strategy alleviates the FRR–TPR trade-off bottleneck caused by information loss during alert aggregation in our experiments. Empirically, a systematic matrix of 200 cases across two testbeds shows that the aggregation bottleneck isstructural, that it persistsunder a different alert-grouping policy, and that Case-Level verification attains operating points unreachable at the Alert Level. From an engineering perspective, independent HTTP replay is a key source of the Case-Level advantage, and testbed-specific response patterns must be adapted when deploying to a new testbed.

## 2 Related Work

### 2.1 Traditional Scanner Evaluation and DAST/SAST

The evaluation of Web vulnerability scanners is a well-established research area. Fonseca et al. were among the first to systematically test and compare the detection and false-positive behavior of multiple Web vulnerability scanning tools for SQL injection and XSS, identifying both substantial differences in detection capability across tools and a common tendency toward high false-positive rates [1]. Antunes and Vieira subsequently studied SQL injection vulnerability detection in Web-service settings and quantified scanner false-negative and false-positive characteristics [9]. In an empirical comparison of commercial and open-source scanners, Amankwah et al. found that false positives were a primary obstacle to practical adoption regardless of scanner type [2]. A systematic literature review by Alazmi and Conte de Leon synthesized the characteristics and effectiveness of Web application vulnerability scanners and likewise identified high false-positive rates as a common bottleneck [4]. These studies focus on the scanners' own detection rates and false-positive behavior, establishing the background for the high-recall, high-false-positive characteristics of conventional scanners considered in this work. They generally do not investigate downstream governance of scanner detections, that is, how false positives can be filtered while preserving true positives.

### 2.2 LLM-Based Vulnerability Detection

As LLM capabilities have advanced, their application to vulnerability detection has attracted considerable attention. Ridoy et al. proposed EnStack, an ensemble stacking framework that combines multiple LLMs to improve source-code vulnerability detection [5]. Lopes et al. examined the ability of open-source LLMs to identify vulnerabilities, reporting substantial recall but limited precision [6]. Isozaki et al. developed an LLM benchmark for automated penetration testing and analyzed opportunities for improvement [7]. For specific vulnerability classes, dedicated machine-learning and deep-learning techniques have been investigated for both SQL injection [10] and XSS detection [11]. Survey studies [8,12] have systematically reviewed the use of LLMs in cybersecurity and threat detection. Most of this work, however, examines the independent detection capability of LLMs. Few studies have systematically quantified LLM performance in the subordinate role of verifying scanner outputs, and none have addressed the structural effect of verification granularity on the upper bound of false-positive filtering. This gap motivates the present study.

### 2.3 Hybrid Detection, Cascades, and False-Positive Mitigation

Combining multiple detectors in a cascade or ensemble is a classic approach to improving overall performance. In vulnerability management, Millar et al. used deep learning to improve DAST vulnerability triage and thereby reduce the burden of manual review [3]. False-positive mitigation for static-analysis alerts has also emerged as a distinct research area. Guo et al. systematically reviewed progress, challenges, and opportunities in mitigating false-positive alerts from static analysis [13]. Murali et al. proposed FuzzSlice, which uses function-level fuzzing to eliminate false-positive static-analysis alerts [14]. Lin introduced an LLM-driven method for false-positive mitigation [15]. Although related to this work, those cascades typically comprise homogeneous feature-vector classifiers or program-analysis components that can be modeled statistically. By contrast, the second stage considered here is a natural-language-prompt-driven LLM whose decisions are highly sensitive to input granularity and prompt wording, making existing ensemble theory difficult to apply directly.

### 2.4 Prompt Engineering, Safety Alignment, and Ablation Methodology

Prompt engineering substantially affects LLM reasoning performance, with Chain-of-Thought (CoT) prompting being a representative technique [16]. Methodologically, conventional ablation studies often remove only one component at a time, implicitly assuming additive component effects and potentially overlooking interactions. We therefore include joint-removal controls as a standard component of the ablation design.

### 2.5 Limitations of Prior Work and Positioning of This Study

Taken together, four gaps remain across these research areas: traditional scanner evaluations focus on scanners themselves while overlooking downstream governance; LLM-based vulnerability detection predominantly examines independent detection rather than the subordinate role of verification; hybrid-cascade research mainly targets homogeneous classifiers rather than prompt-driven LLM verifiers; and safety-alignment and ablation methodologies have not yet been integrated with cascade-verification tasks. The closest points of comparison are LLM-based false-positive mitigation for static analysis [15] and DAST triage [3]. These studies likewise use learning-based methods to filter detector false positives, but emphasize model capability and enhanced detection. The present work instead focuses on verification granularity as a structural variable: without changing the model, merely undoing alert aggregation and shifting decision granularity downward can expand the set of feasible FRR–TPR operating points. This distinction between verification granularity versus model capability and the upper bound of false-positive filtering versus enhanced detection capability defines the central positioning of this work relative to prior studies.

## 3 Methodology

### 3.1 Two-Stage Cascaded Hybrid Detection Architecture

We employ a two-stage cascade consisting of scanner detection followed by LLM verification. In the first stage, Burp

Suite Professional performs a full active scan of the target Web application and outputs candidate detected cases. In the second stage, the HTTP request for each detected case is reconstructed and resent, and the actual response is collected. The request context and response are then supplied to the LLM, which classifies the case as a true positive or false positive. Cases classified as false positives are removed from the final results.

The division of responsibilities within the two-stage cascade is to preserve recall while reducing false positives. The recall ceiling of the cascade cannot exceed that of the scanner itself; the value of the LLM lies in filtering as many false positives as possible while sacrificing as few true positives as possible. A crucial aspect of this formulation is that the LLM acts as a verifier rather than a detector, and its input is always anchored to a specific case already detected by the scanner. This differs fundamentally from standalone LLM detection with a high FPR. The scanner sharply narrows the search space of the verification task, requiring the LLM only to make a binary decision on an existing candidate rather than autonomously discover vulnerabilities across the entire request space. Consequently, the performance bottleneck of the cascade shifts away from the LLM's detection capability and toward the design of the verification workflow, specifically, the granularity at which candidates are organized and the manner in which evidence is presented to the LLM.

### 3.2 Alert-Level and Case-Level Verification Granularity

Burp's native reports follow a three-level Issue → Alert → Instance hierarchy. An Issue represents a vulnerability type (e.g., "SQL injection"). An Alert corresponds to a URL path and constitutes Burp's native aggregation granularity; it is therefore used as the Alert-Level decision unit in this study. An Instance is an indivisible HTTP request–response pair and is used as the Case-Level decision unit. Figure 1 illustrates this three-level mapping. Importantly, Case-Level verification does not introduce a new concept; rather, it strategically exploits information already available at the scanner's Instance layer. Each Instance is inherently an atomic unit in a Burp report. This work simply elevates it from a passive reporting unit to an active verification decision unit.

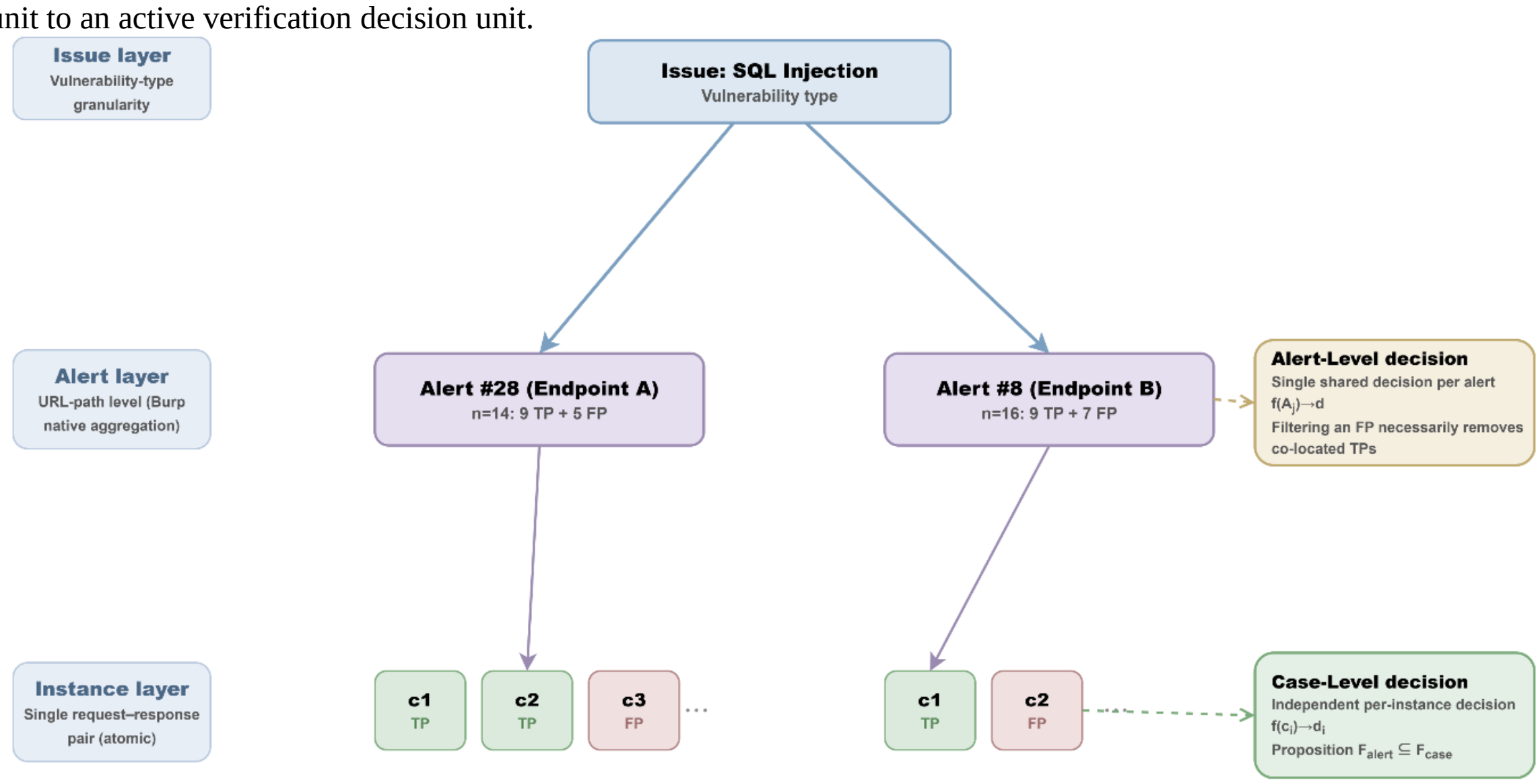

**Figure 1: Issue–Alert–Instance Hierarchical Mapping**

Figure 1 makes the information loss caused by aggregation immediately apparent. Once multiple Instances are collapsed into a single Alert, Alert-Level verification can issue only one decision for the entire Alert, erasing differences between true and false Instances within the group. The mechanism is formalized below.

Definition 1 (Verification Granularity): Let the alert aggregation mapping be $A: \{c_1, \dots, c_{91}\} \longrightarrow \{A_1, \dots, A_{30}\}$, which maps 91 detected cases to approximately 30 alerts. For the $j$-th alert, let the $n_j$ cases aggregated within it be $A_j = \{c_1, c_2, \dots, c_{n_j}\}$.

Alert-Level verification requires the LLM to issue a single decision for the alert as a whole; all cases in the group share that decision:

$$f(A_j) \longrightarrow d \in \{TP, FP\}, d_1 = d_2 = \cdots = d_{n_j} = d \quad (1)$$

Case-Level verification instead adjudicates each case independently, removing the equality constraint on decisions within a group:

$$f(c_i) \longrightarrow d_i \in \{TP, FP\}, i = 1, \dots, n_j \quad (2)$$

This yields the core proposition $F_{alert} \subseteq F_{case}$: the set of feasible Alert-Level operating points is a subset of the set of feasible Case-Level operating points. Any (FRR, TPR) combination attainable at the Alert Level can also be attained at the Case Level by imposing identical decisions within each group, thereby reducing the Case-Level strategy to an Alert-Level strategy; the converse does not hold. Operationally, this result arises because Alert-Level verification enforces a uniform decision within each group, whereas Case-Level verification preserves an independent decision space for every case.

For alert $j$, let $p_j$ denote the number of true positives and $f_j$ the number of false positives. Let $x_j \in \{0,1\}$ indicate whether the entire alert is classified as a false positive. The false-positive reduction rate and true-positive loss can then be written as follows:

$$FRR = \frac{\sum_j x_j f_j}{\sum_j f_j}, TP_loss = \sum_j x_j p_j \quad (3)$$

### 3.3 HTTP Replay Mechanism

The technical prerequisite for Case-Level verification is the ability to resend the HTTP request associated with each Burp-detected case independently and reproducibly, and to collect the actual response. The replay implementation addresses the following engineering details: (1) request extraction: the complete HTTP request is reconstructed by Base64-decoding the request field in the Burp JSON report; (2) replay method: each request is resent from scratch using the Python requests library rather than replayed through Burp Repeater, thereby excluding interference from Burp's internal state; (3) session and authentication: DVWA uses PHPSESSID and WebGoat uses JSESSIONID; the system logs in before execution, obtains and injects the cookie, and automatically reauthenticates upon expiration; (4) CSRF token: because DVWA forms contain a one-time user_token, a two-stage procedure first parses the token from a GET response and then embeds it in the POST body; (5) ordering and concurrency: requests are executed sequentially in a single thread and in ascending request-index order to prevent races; (6) response truncation: the body is truncated to the first 4096 characters to satisfy context limits while retaining the region in which the injection output appears; (7) binary responses: for types such as image/*, only a type placeholder is returned; (8) testbed state reset: database-reset scripts and lesson initialization maintain a clean environment. An uninitialized WebGoat lesson returns a null pointer exception (NPE) response, which is handled explicitly by element E5 of the CEV-Prompt; (9) multi-step attacks: all 91 detections in this experiment are single-step requests; and (10) two-stage replay for stored XSS: the injection and page-loading stages are replayed and adjudicated independently without sharing a decision.

### 3.4 Five Elements of the CEV-Prompt (Plan A)

The primary verification strategy uses a prompt formally designated the Case-Focused Evidence Verification Prompt (CEV-Prompt; hereafter, Plan A). It comprises five independently switchable segments, allowing the corresponding segment to be removed precisely for ablation. Table 1 summarizes the five elements.

Table 1: Five Elements of the CEV-Prompt (Plan A)

| Element | Name | Function |
|---|---|---|
| E1 | Page anatomy guide | Directs attention to key regions of the page structure (e.g., DVWA's main-content container) |
| E2 | Vulnerability-specific | Specifies where evidence should appear for each vulnerability type (e.g., SQL |

| Element | Name | Function |
|---|---|---|
| | evidence localization | injection evidence before the title and command-injection evidence within <pre>) |
| E3 | Anti-hallucination checklist | Requires confirmation that evidence is present, excludes interference from instructional links and navigation menus, and suppresses unsupported inference |
| E4 | Pitfall warning | Warns against misleading echoes such as “More Information” |
| E5 | Anomalous-response pattern handling | Explicitly identifies anomalous responses such as WebGoat NPEs to prevent systematic misclassification |

As a control, we also design an Input Preprocessing Prompt (IPP; hereafter, Plan B), which emphasizes preprocessing of the input information, including response cleaning and structured summarization. This control is used to determine whether the bottleneck lies in the reasoning strategy or the amount of information provided.

### 3.5 Evaluation Metrics

The core metrics are defined as follows: true positive rate, $TPR = TP/(TP + FN)$; false positive rate, $FPR = FP/(FP + TN)$; the principal metric for quantifying the value of the cascade, False-positive Reduction Rate, $FRR = (Burp_FP - Cascade_FP)/Burp_FP$; cascade true positive rate, $Cascade_TPR = Cascade_TP/GT_pos$; and cascade false positive rate, $Cascade_FPR = Cascade_FP/GT_neg$. F1 is the harmonic mean of precision and recall. Annotation agreement is measured using Cohen’s Kappa [17].

To report the true-positive cost explicitly alongside FRR, we further define three derived metrics: scanner-relative true-positive retention, $TP\ Retention$, i.e., the proportion of Burp true positives retained; the true-positive loss rate, $TPLR$; and the FP–TP exchange ratio, defined as the number of true positives lost for every 1 false positive filtered.

## 4 Experimental Design

### 4.1 Dataset and Ground-Truth Annotation

We construct an evaluation set containing 200 cases; Table 2 reports its distribution and annotation agreement. The cases comprise 30 instances each of command injection, file inclusion, SQL injection, and cross-site scripting, together with 80 benign cases, yielding 87 positive and 113 negative cases. Test payloads are drawn from three public repositories (PayloadsAllTheThings, SecLists, and sqlmap’s built-in payload library) and manually screened for compatibility with the testbeds. Obfuscated variants are generated by an LLM through URL encoding, case transformations, comment insertion, whitespace substitution, and quotation-mark variants, after which their actual behavior is manually verified.

Table 2: Dataset Distribution and Annotation Agreement (200 Cases)

| Dimension | Distribution |
|---|---|
| Total cases | 200 (87 positive / 113 negative) |
| Vulnerability type | Benign 80; command injection 30; file inclusion 30; SQL injection 30; cross-site scripting 30 |
| Testbed | DVWA 137, WebGoat 63 |
| Difficulty | Basic 66; obfuscated 54; not applicable 80 |
| Annotation agreement | Agreement 179 / adjudication 21 (disagreement rate 10.5%); Cohen’s Kappa=0.791 |

### 4.2 Testbed Environment

Experiments are conducted on two instructional testbeds, Damn Vulnerable Web Application (DVWA) and WebGoat. Their complementary technology stacks and vulnerability implementations reduce dependence on any single testbed. DVWA (v1.10 Development, Low security level, PHP 7.x + MySQL 5.x) contributes 137 form-level cases across five categories (68.5%). WebGoat (v2023.4, Java 17 + Spring Boot, default security level) contributes 63 cases involving multi-step lesson workflows (31.5%). Both testbeds are deployed through Docker Desktop, using the vulnerables/web-dvwa and webgoat/webgoat images on ports 80 and 8080, respectively. Database state is reset before every experimental run: DVWA by replaying reset.sql and WebGoat by restarting the container. The Low security level is selected to ensure effective payload injection and stable detection patterns.

### 4.3 Burp Suite Professional Scan Configuration

The first-stage scan uses the Active Scan engine in Burp Suite Professional v2023.12.x. Both the audit configuration and crawl depth retain their default settings; authentication is maintained through a preconfigured session cookie; and the scan scope is strictly limited to the testbed IP addresses and ports. The output is a Burp JSON report containing Base64-encoded raw requests and responses, organized according to the Issue → Alert → Instance hierarchy described in Section 3.2. Across the 200 cases, Burp detects 91 cases (70 true positives and 21 false positives), which are aggregated into approximately 30 Alerts. Only about 5 Alerts contain false positives; the 21 false positives coexist with numerous true positives in this small set of Alerts, providing the data-structural basis of the aggregation bottleneck formalized in Section 3.2.

### 4.4 LLMs and API Configuration

The second-stage verification uses 6 commercial LLMs, including GLM-4, GLM-5.2, Qwen-Max, Qwen3.7-Max, DeepSeek-V4-Flash, DeepSeek-V4-Pro. All models use the same invocation parameters: stream=False, Temperature=0 for reproducibility, top_p=1.0, and max_tokens=512.

### 4.5 Comparison Strategies and Experimental Matrix

To systematically disentangle the contributions of granularity, prompt design, and model choice, we define 18 comparison strategies (S0–S17), as shown in Table 3. The matrix follows a progressive design: S0–S2 measure the empirically best operating point under the Alert-Level aggregation constraint; S3–S11 and S12 evaluate the granularity effect, the main approach, and ablations; and S13–S17 provide additional controls for leave-one-testbed-out validation, controlled granularity, no-replay conditions, and a deterministic rule-based baseline. In total, the experiments comprise 31 independent runs and approximately 4900 API calls.

Table 3: Comparison Strategy Matrix (S0–S17)

| ID | Strategy | Granularity | Model | Prompt Variant | Purpose |
|---|---|---|---|---|---|
| S0 | Alert Zero-shot | Alert | G4/QM/DS-V4-Flash | zero-shot | Ceiling baseline |
| S1 | Alert Few-shot | Alert | All 6 models | few-shot | Ceiling baseline |
| S2 | Alert CoT | Alert | G4/QM/DS-V4-Flash | CoT | Ceiling baseline |
| S3 | Case Baseline | Case | All 6 models | case-fewshot | Granularity effect |
| S4 | Case Plan A | Case | All 6 models | case-focused | Primary approach |
| S5 | Case Plan B | Case | G4/G52/Q37/DS-V4-Pro | preprocessed | Preprocessing control |

| ID | Strategy | Granularity | Model | Prompt Variant | Purpose |
|---|---|---|---|---|---|
| S6 | −E1 (without page anatomy) | Case | GLM-4 | case-focused−E1 | Single-element ablation |
| S7 | −E2 (without evidence localization) | Case | GLM-4 | case-focused−E2 | Single-element ablation |
| S8 | −E3 (without anti-hallucination guidance) | Case | GLM-4 | case-focused−E3 | Single-element ablation |
| S9 | −E4 (without pitfall warning) | Case | GLM-4 | case-focused−E4 | Single-element ablation |
| S10 | −E5 (without anomalous-response handling) | Case | GLM-4 | case-focused−E5 | Single-element ablation |
| S11 | E3+E5 only | Case | G4/QM/DS-V4-Flash | e3e5-only | Joint-ablation control |
| S12 | Plan A + Plan B | Case | GLM-4 | focused + preprocessed | Combination control |
| S13 | E1a DVWA→WebGoat leave-one-testbed-out | Case | GLM-4 | case-focused (without E5) | Leave-one-testbed-out validation |
| S14 | E1b WebGoat→DVWA leave-one-testbed-out | Case | GLM-4 | case-focused | Leave-one-testbed-out validation |
| S15 | E2 controlled-granularity comparison | Hybrid | GLM-4 | controlled-granularity | Exclude prompt confounding |
| S16 | E6 no-replay baseline | Case | GLM-4 | case-fewshot (without replay) | Quantify the contribution of replay |
| S17 | E7 deterministic rule-based baseline | Case | — | Python regular expressions | Upper bound of filtering without an LLM |

## 5 Experimental Results

### 5.1 Standalone Detection Baseline

Table 4 reports the standalone detection performance of 7 tools/models across the 200 cases.

Table 4: Standalone Detection Baseline (7 Tools/Models × 200 Cases)

| Tool/Model | Detected | TPR | FPR |
|---|---|---|---|
| Burp Suite Pro | 91 | 80.5% (70/87) | 18.6% (21/113) |
| GLM-4 | 183 | 100.0% (87/87) | 85.0% (96/113) |
| Qwen-Max | 163 | 100.0% (87/87) | 67.3% (76/113) |
| DeepSeek-V4-Flash | 158 | 100.0% (87/87) | 62.8% (71/113) |
| GLM-5.2 | 154 | 100.0% (87/87) | 59.3% (67/113) |

| Tool/Model | Detected | TPR | FPR |
|---|---|---|---|
| Qwen3.7-Max | 148 | 95.4% (83/87) | 57.5% (65/113) |
| DeepSeek-V4-Pro | 140 | 96.6% (84/87) | 49.6% (56/113) |

As shown in Table 4, the 6 LLMs achieve standalone detection recall of 95.4%–100%, but their false positive rates are likewise high, ranging from 49.6% to 85.0%. These excessive false positive rates preclude LLMs from independently replacing conventional scanners. Burp Suite Professional achieves a TPR of 80.5% and an FPR of 18.6%, demonstrating high recall with a moderately high false positive rate and making it suitable as the first-stage detector in the cascade.

### 5.2 The Alert-Level Trade-off Bottleneck

Cascade verification is performed at Alert-Level granularity across 6 models and 3 strategies (Zero-shot, Few-shot, and CoT), with the results shown in Table 5. The central finding is that none of the configurations achieves an FRR above 42.86%. This empirically best operating point is highly consistent across the 6 tested models and 3 strategies and corresponds to a cascade TPR of 51.7%.

Table 5: Alert-Level Cascade Results

| Model | Strategy | Confirmed TPs | Filtered FPs | Cascade TPR | Cascade FPR | FRR |
|---|---|---|---|---|---|---|
| GLM-4 | Zero-shot | 29/30 | 1/30 | 0.805 | 0.186 | 0.0% |
| GLM-4 | Few-shot | 22/30 | 8/30 | 0.701 | — | 23.8% |
| Qwen-Max | Zero-shot | 10/30 | 20/30 | 0.494 | 0.106 | 42.9% |
| Qwen-Max | Few-shot | 22/30 | 8/30 | 0.701 | — | 23.8% |
| Qwen-Max | CoT | 22/30 | 7/30 | 0.517 | 0.106 | 42.9% |
| DeepSeek-V4-Flash | Few-shot | 21/30 | 9/30 | 0.517 | 0.106 | 42.9% |
| DeepSeek-V4-Flash | CoT | 17/30 | 13/30 | 0.517 | 0.106 | 42.9% |
| GLM-5.2 | Few-shot | 23/30 | 7/30 | 0.517 | 0.106 | 42.9% |
| Qwen3.7-Max | Few-shot | 27/30 | 3/30 | 0.517 | 0.106 | 42.9% |
| DeepSeek-V4-Pro | Few-shot | 24/30 | 6/30 | 0.517 | 0.106 | 42.9% |

Note: The denominator of FRR is all 21 false positives produced by Burp. A total of 12 combinations were executed. Two additional configurations, GLM-4 CoT (FRR=0.0%) and DeepSeek-V4-Flash Zero-shot (FRR=23.8%), are omitted from the table because both FRRs fall below the empirically best operating point of 42.86% and do not affect the conclusion.

The 42.86% value is not an upper bound reached fortuitously by a particular model; rather, under the alert-aggregation constraint, it is the empirically best operating point determined by the data structure itself. Because the 21 false positives are collapsed into 5 alert endpoints that also contain true positives, any Alert-Level strategy that filters those false positives must remove the corresponding alerts in their entirety, thereby eliminating the true positives within the same groups. Changing the strategy (for example, from Zero-shot to CoT) can therefore move the operating point only along this bottleneck curve, not beyond it. GLM-4 Zero-shot, for example, is so permissive that it filters almost no false positives (FRR=0), whereas more aggressive strategies filter additional false positives at the cost of greater true-positive loss. This observation locates the bottleneck in the aggregated data structure rather than in model capability, directly motivating the shift in granularity examined in the next section. Importantly, 42.86% is the empirically best operating point within the scope of this experiment, not a theoretical upper bound; the value itself depends on the specific alert topology, as analyzed in Section 6.1.

### 5.3 Effect of Case-Level Granularity

After shifting verification granularity from the Alert level to the Case level, the results of the complete GLM-4 strategy set are shown in Table 6. The granularity effect is first evident in the Baseline: the Case-Level Baseline achieves FRR=47.6%, an improvement of 4.7 percentage points (+4.7 pp) over the empirically best Alert-Level operating point of 42.86%, while cascade TPR increases from 51.7% to 55.2%. This result supports the proposition in Section 3.2, $F_{alert} \subseteq F_{case}$: after aggregation is removed, per-case decisions can independently filter false positives that were previously bound to other cases.

Table 6: Case-Level Cascade Results (GLM-4)

| Plan | Strategy | Cascade TPR | Cascade FPR | FRR | LLM Accuracy | F1 |
|---|---|---|---|---|---|---|
| Burp (baseline) | — | 0.805 (70/87) | 0.186 (21/113) | — | — | — |
| Case-Level Baseline | S3 | 0.552 (48/87) | 0.097 | 47.6% (10/21) | 0.637 (58/91) | 0.657 |
| Plan A (CEV-Prompt) | S4 | 0.621 (54/87) | 0.097 | 47.6% (10/21) | 0.703 (64/91) | 0.711 |
| Plan B (preprocessing) | S5 | 0.552 (48/87) | 0.097 | 47.6% (10/21) | 0.637 (58/91) | 0.657 |
| Plan A+B | S12 | 0.621 (54/87) | 0.097 | 47.6% (10/21) | 0.703 (64/91) | 0.711 |

The value of Plan A, the Case-Focused Evidence Verification Prompt, can be stated precisely: at exactly the same FRR of 47.6% as the Baseline, it increases cascade TPR from 55.2% to 62.1%. In other words, Plan A does not filter more false positives; it preserves more true positives while filtering the same number of false positives. The confusion matrices in Figure 2 substantiate this mechanism.

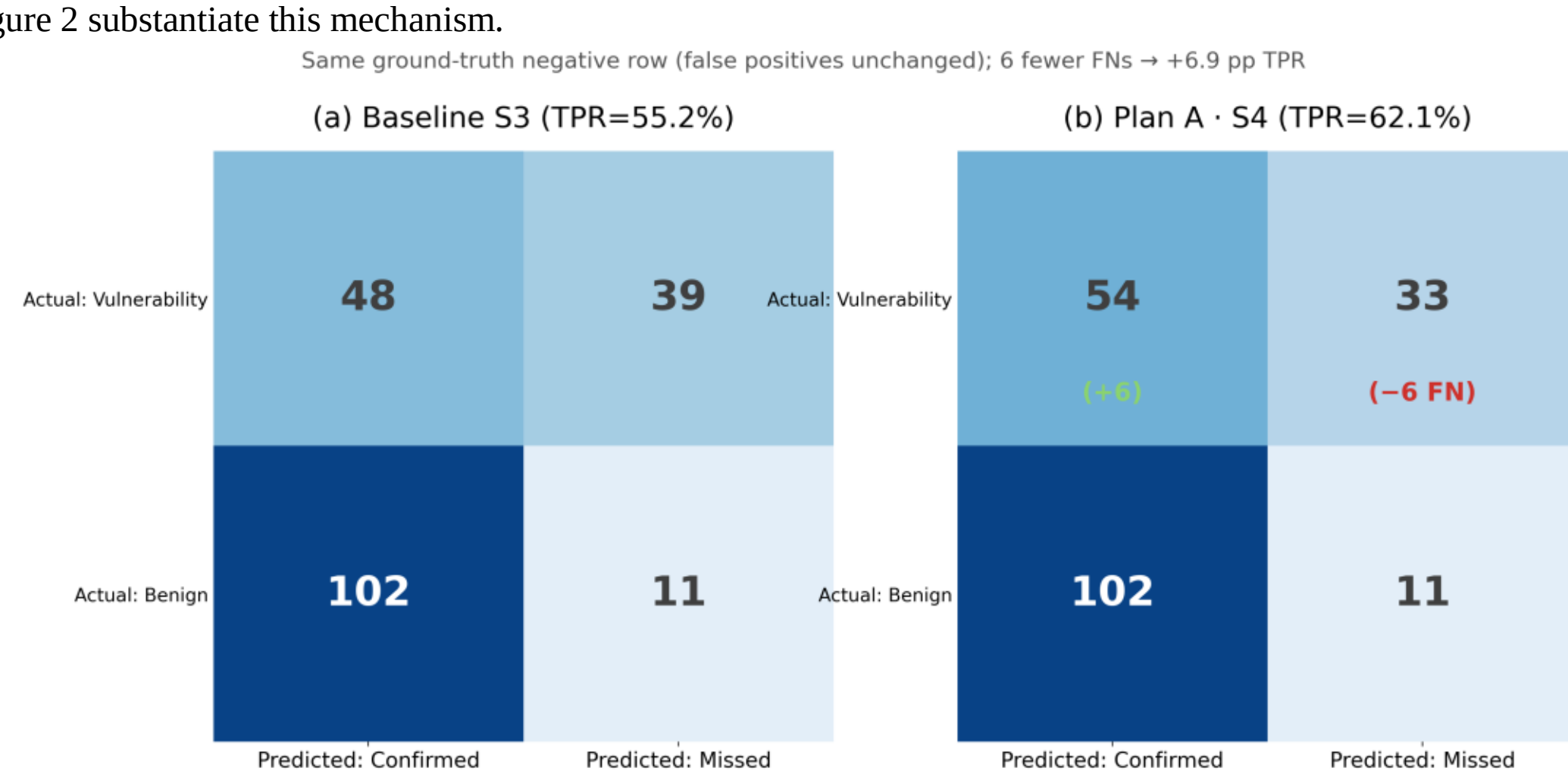

**Figure 2: Comparison of Confusion Matrices**

Comparison of the left and right panels in Figure 2 (Baseline and Plan A, respectively) yields three observations. First, both strategies have exactly the same FRR (10/21), and their ground-truth negative rows are identical: 11 false positives are classified as true positives, while 102 cases are correctly filtered. Thus, Plan A has no additional effect on the classification of benign cases. Second, in the ground-truth positive row, Plan A recovers 6 true positives that the Baseline misclassified as false positives (39→33); this is the sole source of the +6.9 pp increase in TPR. Third, because the ground-truth negative row remains unchanged, Plan A filters no additional false positives; its value is strictly confined to true-positive preservation. By contrast, Plan B (the Input Preprocessing Prompt) produces exactly the same metrics as the corresponding baseline whether used alone or combined with Plan A, indicating that the bottleneck lies in the reasoning strategy rather than the amount of input information. The same operating points can also be examined as ROC anchors, as shown in Figure 3.

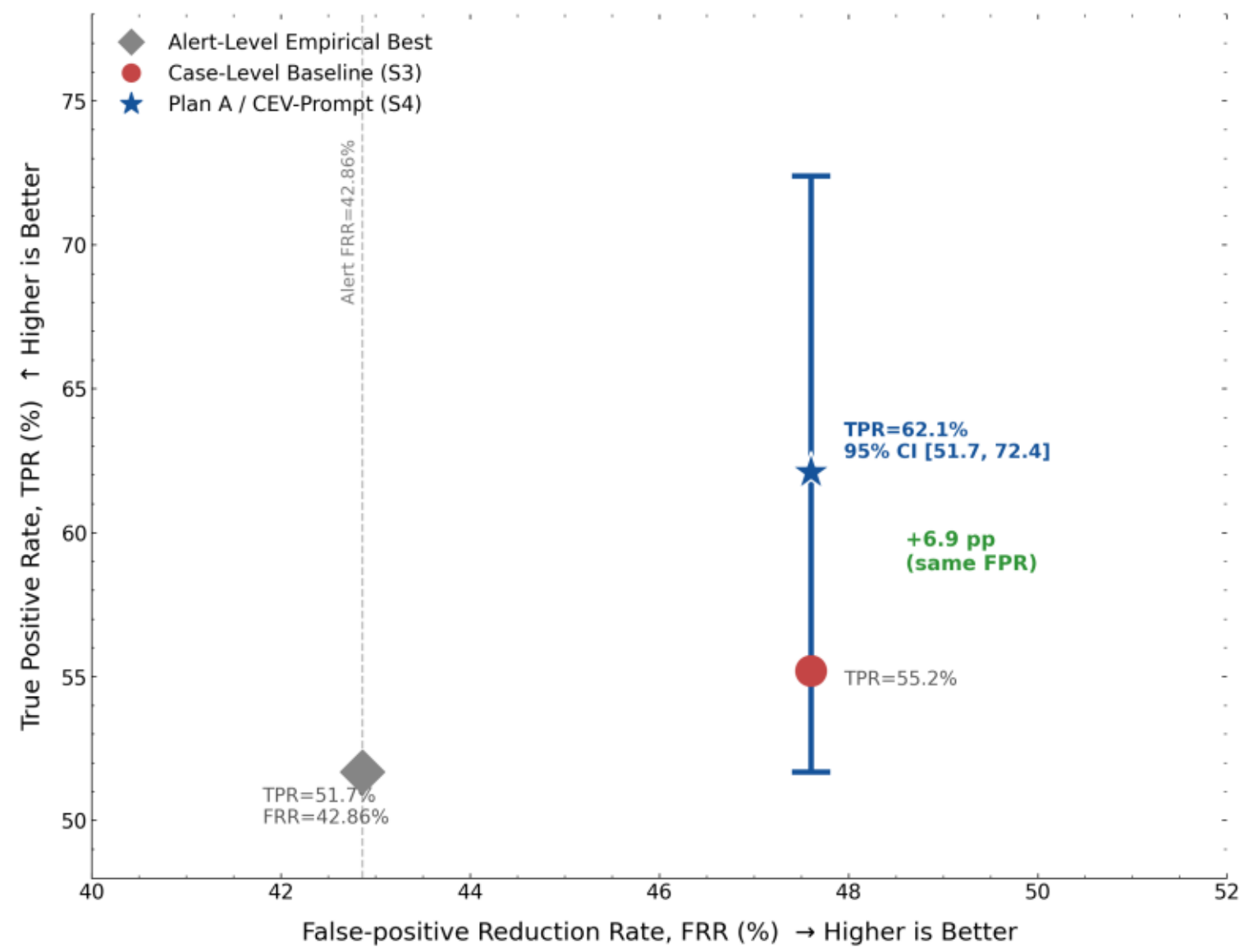

**Figure 3: Comparison of FRR–TPR Operating Points**

Figure 3 plots the Alert-Level and Case-Level operating points with bootstrap 95% confidence intervals. At the same false-positive-rate anchor (FPR≈0.09), the Case-Level strategy provides higher recall than the Alert-Level strategy. At the false-positive level corresponding to the empirically best Alert-Level operating point (FRR=42.86%), the Case-Level operating point has already surpassed it. This result rules out the alternative explanation that the Case-Level advantage merely reflects a lower decision threshold that trades more false positives for higher recall. Instead, Case-Level verification achieves higher recall at the same or a lower false positive rate, representing a genuine expansion of the feasible operating-point set rather than a translation along the same curve. Taken together, the evidence in this section shows that shifting granularity downward expands the feasible FRR–TPR operating-point set in our data, answering RQ2, while the CEV-Prompt serves to preserve true positives within that set. For the derived metrics, Plan A achieves scanner-relative true-positive retention of 77.1% (54/70), a true-positive loss rate of 22.9%, and an FP–TP exchange ratio of approximately 1.6:1.

### 5.4 Ablation Study

To identify the contribution of each of the five CEV-Prompt elements, we remove each element individually under GLM-4 and observe the resulting change in FRR. Table 7 presents the results.

Table 7: Ablation Study (GLM-4, Five Elements of Plan A)

| Strategy | FRR | TPR | ΔFRR vs Full | ΔTPR vs Full | Interpretation |
|---|---|---|---|---|---|
| Full Plan A | 47.6% | 62.1% | Reference | Reference | Reference |
| −E5 (anomalous-response pattern handling) | 23.8% | 67.8% | −23.8pp | +5.7pp | Effective for FRR |
| −E3 (anti-hallucination guidance) | 33.3% | 64.4% | −14.3pp | +2.3pp | Effective for FRR |
| −E1 (page anatomy) | 52.4% | 57.5% | +4.8pp | −4.6pp | Effective for TPR |
| −E2 (evidence localization) | 52.4% | 59.8% | +4.8pp | −2.3pp | Effective for TPR |
| −E4 (pitfall warning) | 52.4% | 56.3% | +4.8pp | −5.8pp | Effective for TPR |
| E3+E5 only (−E1−E2−E4) | 33.3% | 66.7% | −14.3pp | +4.6pp | Protective interaction |

Removing E5 reduces FRR to 23.8%, while removing E3 reduces it to 33.3%, indicating that these elements normally promote false-positive filtering; E5 and E3 are therefore the two elements that are effective for FRR. By contrast, removing any of E1, E2, or E4 increases FRR to 52.4%, indicating that these elements normally restrain false-positive filtering and thus exert a negative effect on FRR. Inspection shows that removal reverses the model's erroneous decisions on borderline cases such as C0054. Moreover, retaining only E3+E5 yields FRR=33.3%, exactly matching the result obtained by removing E3 alone. This reveals a protective interaction: in the absence of E1/E2/E4, E3 still preserves the core filtering capability of E5. Specifically, the 3 additional unfilterable false positives introduced by removing E5 (C0034, C0039, and C0042) are suppressed when E3 is present. E3 therefore acts as a safety net for E5. This finding has methodological implications: conventional single-element ablation under an additivity assumption would mischaracterize the negative direction of E1/E2/E4 and overlook the nonadditive E3–E5 interaction. Joint-removal controls should therefore be a standard component of prompt ablation studies. The negative effects observed here are specific to this experimental setting and arise from decision reversals on particular borderline cases; they should not be generalized as universal properties of the element designs.

From the joint FRR–TPR perspective, removing E3 or E5 lowers FRR but raises TPR, whereas removing E1, E2, or E4 raises FRR but lowers TPR. Removing E1/E2/E4 filters 1 additional false positive at the cost of incorrectly eliminating 2 to 5 true positives. The filtering capability of E3/E5 likewise comes at a cost to TPR. No variant outperforms Plan A on both metrics; the ablations therefore do not constitute alternatives to the primary strategy, but instead show that the five elements move the operating point in different directions along the trade-off curve.

### 5.5 Breakdown by Vulnerability Type

Decomposing Plan A's filtering behavior by vulnerability type reveals strong category dependence, as shown in Table 8. All 10 false positives filtered by Plan A are SQL injection cases (10/12, FRR=83.3%), but the true-positive retention for that type is only 27.8% (5/18). True and false SQL injection cases are highly similar at the level of HTTP response text, so the high FRR is achieved at the cost of substantial true-positive loss. By contrast, command injection and file inclusion both retain 100% of their true positives, but no false positives are filtered (FRR=0%). The 4 XSS false positives remain unfilterable under both the complete Plan A configuration and all ablation variants (FRR=0%), while 81.3% of XSS true positives are retained (13/16).

Table 8: Breakdown by Vulnerability Type (GLM-4 Case-Level Plan A)

| Type | Burp FP | Plan A FRR | Burp TP | TPs Retained by Plan A |
|---|---|---|---|---|
| SQL injection | 12 | 83.3% (10/12) | 18 | 27.8% (5/18) |
| Cross-site scripting (XSS) | 4 | 0.0% (0/4) | 16 | 81.3% (13/16) |
| Command injection | 4 | 0.0% (0/4) | 18 | 100.0% (18/18) |
| File inclusion | 1 | 0.0% (0/1) | 18 | 100.0% (18/18) |

Note: The denominator of FRR is the number of Burp false positives for each vulnerability type, and the denominator of TP retention is the number of Burp true positives for that type. The 4 XSS false positives cannot be filtered through black-box HTTP verification and require dynamic browser-based validation.

The differences across vulnerability types reflect an inherent limitation of the method rather than a limitation of model capability. XSS false positives cannot be filtered at the pure HTTP layer because true and false echoes are indistinguishable in text; dynamic browser rendering is required to determine whether a script actually executes. The high FRR for SQL injection, meanwhile, arises from database error echoes. Such echoes may result either from a genuine injection or from a benign payload containing an SQL fragment, and their textual similarity makes true and false cases difficult to distinguish during filtering. This breakdown directly supports the applicability of Plan A: it is effective for command injection and file inclusion, requires a trade-off for SQL injection, and is unsuitable for XSS.

### 5.6 Statistical Tests

We estimate intervals for Plan A's core metrics using bootstrap resampling[18] (5000 resamples, seed=42, percentile method). The TPR point estimate is 62.07%, with a 95% confidence interval of [51.72%, 72.41%]; the FRR point estimate is 47.62%, with a 95% confidence interval of [19.05%, 76.19%]. The extremely wide FRR interval directly reflects the insufficient sample of only 21 filterable false positives. McNemar's exact test[19,20], used to compare the paired differences between the two classifiers, yields 13 wins and 7 losses for Plan A relative to the Baseline, with a two-sided p=0.2632. The direction is favorable but not statistically significant. Power analysis indicates that approximately 200 false positives would be required to detect a 10 pp improvement in FRR with 80% power; the current sample of 21 is far from sufficient. For reproducibility, 3 independent runs at Temperature=0 yield a TPR standard deviation of $\sigma \leq 0.6$ pp and an FRR standard deviation of $\sigma = 0$ pp, ruling out API randomness as a material factor. Accordingly, all claims about Plan A are strictly limited to directional evidence and do not assert statistical significance or generalizability.

### 5.7 Auxiliary Control Experiments

Three auxiliary control experiments further isolate the source of the Case-Level advantage and rule out competing hypotheses. E2, the controlled-granularity comparison, applies the same Case-Level few-shot prompt under GLM-4 while changing only the decision granularity. Case-Level verification continues to outperform Alert-Level verification, excluding the hypothesis that the advantage is caused by prompt differences and confirming granularity itself as a causal factor. E6, the no-replay baseline, achieves FRR=33.3% when decisions are based on Burp's cached responses, compared with 47.6% under independent replay, a difference of −14.3 pp. Decisions differ for 8 cases, with 4 true-positive and 4 false-positive reversals, demonstrating that independent HTTP replay is a key engineering source of the Case-Level improvement in FRR. E7, the deterministic rule-based baseline, uses Python regular expressions to classify all 52 detections covered by the rules as false positives, resulting in TPR=0%. Keyword matching cannot distinguish true from false positives because a database error echo may arise either from a genuine injection or from a benign payload containing SQL fragments. This result provides counterevidence that confirms the indispensable role of the LLM's semantic understanding.

### 5.8 Leave-One-Testbed-Out Validation

To evaluate the testbed dependence of E5, we conduct two leave-one-testbed-out experiments, with the results shown in

Table 9. E1a (DVWA→WebGoat, with E5 omitted at test time) yields FRR=0% on the WebGoat subset; without E5, none of WebGoat's false positives can be filtered. E1b (WebGoat→DVWA, with E5 included) yields FRR=37.5% on the DVWA subset. True-positive identification remains robust in both subsets, with a subset TPR of 100% for E1a and 90.2% for E1b.

Table 9: Leave-One-Testbed-Out Validation

| Experiment | Development → Test | Test Detections | Subset TPR | Subset FRR | Overall TPR | Overall FRR |
|---|---|---|---|---|---|---|
| E1a | DVWA→WebGoat (without E5) | 14 (9TP,5FP) | 100.0% (9/9) | 0.0% (0/5) | 10.3% | 0.0% |
| E1b | WebGoat→DVWA (with E5) | 77 (61TP,16FP) | 90.2% (55/61) | 37.5% (6/16) | 63.2% | 28.6% |

Note: Subset TPR/FRR uses the detections in the corresponding testbed as the denominator; the overall columns use the 87 ground-truth positive cases and 21 false positives, respectively.

These results show that E5 is testbed-specific: it is designed specifically for WebGoat's NPE response pattern. When transferred across testbeds, false-positive filtering capability declines substantially, whereas true-positive identification remains unaffected. In practice, deployment to a new testbed should prioritize adaptation of testbed-specific anomalous-response handling rules rather than redesigning the entire prompt. This testbed specificity is also an important component of the external validity of the reported FRR estimates.

### 5.9 Sample-Sensitivity Analysis

Because FRR is computed over only 21 filterable false positives, individual cases can have a disproportionate effect on the reported percentages. To quantify this sensitivity explicitly, we perform a jackknife analysis: each of the 21 false positives is removed in turn, and FRR is recomputed on the remaining 20 cases. The resulting FRR ranges from 45.0% (9/20, when the removed case was itself filtered) to 50.0% (10/20, when it was not), a bandwidth of 5.0 percentage points around the full-samplevalue of 47.6% (10/21); the maximum deviationcaused by any single case is ±2.6 pp. These ranges confirm that the sample sensitivity of FRR is real, and all FRR comparisons in this paper should be read as directional evidence rather than as point estimates of statistical precision. The structural conclusions of this paper, that alert aggregation creates an FRR–TPR trade-off bottleneck and that Case-Level verification expands the feasible operating-point set do not depend on these specific percentages; only the location of the empirically best operating point does.

### 5.10 Cross-Scanner Validation with OWASP ZAP

To examine whether the aggregation bottleneck and the Case-Level advantage persist under a different deduplication and alert-grouping policy, we run OWASP ZAP against the same two testbeds with the same case-level seeding procedure. ZAP detects 82 of the 200 cases (55 true positives and 27 false positives): TPR = 63.2% (55/87), and FPR = 23.9% (27/113).

ZAP emits alerts at plugin-invocation granularity and deduplicates by URL and parameter signatures, so a single case can be covered by multiple alert groups. The 27 false positives are strongly entangled with true positives: 26 of them have instances in seven mixed groups that also contain 4–19 true positives each. Under the lossless constraint, an Alert-Level strategy on ZAP cannot filter any of the 27 false positives without sacrificing true positives in the same group or discarding verifiable evidence: its FRR is structurally capped at 0. The bottleneck thuspersists under a different aggregation policy; what changes is its location and severity, both again explained by the topology of false positives across decision units.

We then apply the same Case-Level verification pipeline to ZAP's 82 detections. ZAP-side TPR is 87.3% (48/55): the seven killed true positives are all command-injection cases whose replayed responses contain no visible command output, for which the LLM cannot observe execution evidence. All eight filtered false positives have instances in the seven mixed groups, so they are removed exactly where Alert-Level verification cannot act, while the pure-group false positives are retained by verification on genuine replay evidence. Any Alert-Level filter that raised FRR toward the case-level value would sacrifice 4–19 true positives per group, collapsing TPR to as low as 65.5% (36/55); the operating point (FRR 29.6%, TPR 87.3%) is therefore attainable only at Case Level.

## 6 Discussion

### 6.1 Dependence of the Empirically Best Operating Point

Arithmetically, the empirically best Alert-Level operating point of 42.86% is 9/21: at most 9 of the 21 false positives can be filtered without excessive true-positive loss. Its value depends on three factors: the topology of false positives across alert endpoints (the aggregation topology); the proportion of true and false cases within each false-positive-containing alert (the degree of entanglement); and the selection of the representative request, that is, which request–response pair within an alert is used by the LLM to decide the entire alert. Selecting a different representative request may change the decision for the whole alert. Counterfactual analysis illustrates this sample dependence. If all 21 false positives were isolated in alert endpoints containing no true positives, FRR could theoretically approach 100%; if all were inseparably grouped with true positives, FRR would approach 0%. The observed 42.86% is a sample-dependent value between these extremes. This means that the value 42.86% itself is sample-dependent: changing the case set or scanner would change the location of the best operating point. We consequently distinguish two levels of conclusion. The existence of the bottleneck is structural and generalizable: whenever a scanner aggregates multiple requests into a single alert, true and false cases necessarily share a decision unit. The specific location of the bottleneck (42.86%), however, is sample-dependent and nontransferable.

### 6.2 Cross-Scanner Generalization of the Aggregation Bottleneck

Aggregation policies differ across DAST scanners. Burp Suite Professional organizes detections into a three-level Issue → Alert → Instance hierarchy in which the Alert level groups instances by URL path. OWASP ZAP instead emits alerts at the level of individual plugin invocations, deduplicating results by a combination of URL and parameter signatures. Commercial scanners such as Acunetix and AppScan typically group detections by vulnerability type across the whole target. These policies differ in the size of the decision unit: URL-level grouping can bind many instances into a single alert, whereas plugin-invocation grouping tends to produce smaller and more homogeneous groups.

Crucially, the central result of this paper does not depend on which policy is used. The Alert-Level point set is a subset of the Case-Level point set whenever multiple instances are collapsed into a shared decision unit, regardless of how groups are formed. Aggregation thus affects the magnitude of the bottleneck, but not whether the bottleneck exists.

### 6.3 Limitations

This study has two limitations. First, LLM memorization cannot be excluded completely because DVWA and WebGoat are public instructional testbeds whose patterns may have appeared in model pretraining. Second, cross-border data transfer and privacy compliance require consideration. Although the local Docker testbeds contain no real user data, HTTP requests in production may include session tokens and personal information. Sending such data to third-party LLM APIsrequires a compliance assessment; locally hostedopen-source models or private deployments offer potential solutions.

## 7 Conclusion and Future Work

This work systematically formalizes and empirically studies the long-overlooked verification-granularity issue in a two-stage cascade of scanner detection followed by LLM verification. Methodologically, we formalize the FRR–TPR trade-off bottleneck caused by alert aggregation and design a Case-Level, per-case verification strategy. Empirically, a systematic matrix of 200 cases, two testbeds, 6 models, 18 strategies, 31 experiments, and approximately 4900 API calls yields one finding: the existence of the alert aggregation bottleneck is structural, whereas Case-Level verification expands the FRR–TPR trade-off space.

Future research can proceed in three directions. First, the false-positive sample should be expanded using a larger collection of testbeds or scan results from real applications to obtain confirmatory statistical evidence. Second, the underlying model should be fine-tuned specifically for Case-Level verification to further improve filtering accuracy. Third, XSS validation should be upgraded from pure HTTP black-box verification to a dual-channel method combining HTTP text with dynamic execution, thereby overcoming the inherent inability to filter XSS false positives observed in this experiment.